# Microhydration effect on structural, energetic and light scattering properties of first branched interstellar molecule ( i-PrCN)

Sumana Chakraborty[1], Swati Routh[2], Madhu Krishnappa[2]

1: Bose Institute, Kolkata, India
2: CPGS, Jain University, Bangalore, India

Abstract:

In this work, we have focused on microsolvation of isopropyl cyanide (iPrCN) as iso propyl cyanide has been recently detected in interstellar space and is of great importance from the astrochemical and bio-chemical point of view for its branching carbon chains. Such branches are needed for many molecules crucial to life, such as the amino acids that build proteins.The phenomenon of the formation of hydrogen bond affects structure, energetic and electric properties of microhydrated isopropyl cyanide [(i-PrCN-(H2O)n) (n=1-3)] and this has been explored by using three different quantum chemical models (i) HF method exploiting Pople's standard 6-31++G(d,p) basis (ii) MP2/6-31++G(d,p) and (iii) DFT B3LYP/6-31++G(d,p). ). It is observed that the structural parameters calculated by the three models display similarities, however model dependence is evident from equilibrium electronic energies of the clusters .Presence of water molecule has a significant effect on the values of dipole moments and polarizabilities. Rayleigh intensities which are calculated using mean polarizibility and polarizibility anisotropy are increased much due to the formation of hydrogen bonding. For CN stretching vibration of isopropyl cyanide, intensification of Raman scattering activities are observed upon complexation.

## Introduction:

Detection of interstellar molecules and studying their origins , properties and interactionshave long been an important aspect of the astrochemical research which relates itself to other research fields in astronomy like astrophysics and astrobiology through the investigations of these astromolecules. The development of highly sophisticated radio-astronomical instruments has helped to detect over 150 molecules, ranging from simple diatomics like CO to more complex organic molecules like ethylene glycol (OCH2-CH2OH), in the interstellar and circumstellar spaces till date and they are of significant biochemical and astrochemical interest.

Most organic molecules present in the interstellar space are just straight chains of carbon atoms. The recently discovered molecule, isopropyl cyanide, on the other hand, has branching carbon chains [2] as it requires the addition of a functional group to a non-terminalcarbon in the chain. Such branches are needed for many molecules which are crucial to life, such as the amino acids that build proteins.The discovery is in agreement with the theory that life's building blocks may have originated in interstellar space.

The hydrogen bond (H-bond), present in different physiochemical and biological processes and reactions is one of the most important inter-molecular interactions which are essential for providing and sustaining life.It plays a key rolein the formation of stable conformations of biomolecular structures like polypeptides, proteins and clusters . [3-11] Various spectroscopic methods like

Infrared (IR) and Raman spectroscopic methods have been developed in past decades, which are used to study the nature of the hydrogen bonds in molecular systems.Classical light scattering or Rayleigh scattering can be also used to study the hydrogen bonding [12-14]. However, investigations of the scattering of light by the H-bonded molecular systems which are of astronomical interest are quite limited,though they could provide information of both microscopic and collective propertiesabout the molecules. Rayleigh light scattering is derived from dipole polarisability which is the basic property that describes the interaction of a molecule with an electric field.

Of late, microsolvation technique based analysis of hydrogen-bond formation exploiting quantum chemical methods has been accepted to reasonable extent among scientific community [16–26]. A small number of solvent molecules are placed strategically in microsolvation, around the hydrogen bondforming sites. Interactions arising out of hydrogen bonds with solvent can trigger large variations in the chemical and structural properties of the solute molecule. Microhydration (microsolvation with water molecules) of biomolecules is of special significancedue to water being the natural medium for biological molecules and impacting their molecular structure and chemical properties to a remarkable degree. Microsolvationis evident in many gas phase processes where water molecules influence the reactivity and structure of the biomolecules via electrostatic interactions. Thus, it is crucial to characterize the microhydration process of biomolecules in solution and the gas phase.

The present work elaborates the effect of H-bond formation on the structure,energetics and electric properties of Isopropyl cyanide and its complexes with water molecules. Elastic light scattering or Rayleigh scattering helps analyse the variations of the electric properties from various conformations of H-bonded i-PrCNsystem. Anisotropy of the electronic polarisationof a molecular systemare related to cross-section and depolarisation of light scattering. H-bond formation which reflects change in the anisotropy, leads to a corresponding variance in the Rayleigh intensities and depolarisation ratios.

An analysis of Rayleigh scattering properties has been made for several hydrogen bonded dimers, trimmers, tetramer of i-PrCN like i-C3H7CN···H2O, i-C3H7CN···2H2O, i-C3H7CN···3H2O with an objective to see the effect of hydrogen bond formation on the molecular interaction with radiation. Here we used Hartreefock method, density function theory, and MP2 theory with basis set 631+G (d, p) to calculate the parameters. We also calculated the rotational constants and the binding energies .

**Computational methods**

The calculations for geometry optimization in this study are performed with three different quantum chemical models: (1)Hartree-Fock method employing Pople's standard 6-31++G(d,p) basis set: HF/6-31++G(d,p) and including electron correlation at the (2)second-order Moller–Plesset perturbation theory(MP2) with the same basis set: MP2/6-31++G(d,p) and (3) density functional theory (DFT) with B3LYP functional (three parameter hybrid exchange functional of Becke with the Lee–Yang– Parr correlation functional) [49,50] with the same basis set:B3LYP/6-31++G(d,p). The basis set used is diffused and polarized split valence type and is generally accepted as sufficient to predict fairly accurate structural and spectroscopic parameters of medium sized molecule. The optimization of isolated i-PrCN molecule as well as that of all i-PrCN··· H2O clusters (i-C3H7CN···(H2O)$_n$ (n=1,2,3)) is further supported by presence of all real frequencies.

The hydrogen bond interaction in the complexes is investigated in terms of binding energy without and with basis set superposition error (BSSE) corrections, which are denoted as $\Delta E$ and $\Delta E^{CP}$ respectively. The binding energies are calculated using the supermolecule approach. Thus for i-PrCN···(H2O)$_n$ (n=1,2,3)):

$$\Delta E = E_{i-\mathrm{Pr}CN \cdots nH2O} - (E_{i-\mathrm{Pr}CN} + nE_{H2O}), n = 1,2,3$$

Where E denotes electronic energies of respective species. BSSE refers to vacant orbital on one atomic centre being used to make up for a basis set deficiency on a neighbouring atom. The correction for the BSSE is considered, in this work, following a modified version of the counterpoise correction method prescribed by Turi and Dannenberg [15]: The binding energies are further corrected with zero-point vibrational energy ( $\Delta E^{CP+ZPE}$ ) from the harmonic frequency analysis.

The finite field approximation is used to compute dipole moments and dipole polarizibilities. The mean dipole polarizability, $\overline{\alpha}$ and the polarizability anisotropy, ( $\Delta\alpha$ )are the invariants of the polarizability tensor and are given by

$$\overline{\alpha} = \tfrac{1}{3}(\alpha_{xx} + \alpha_{yy} + \alpha_{zz})$$

and

$$(\Delta\alpha)^2 = \tfrac{1}{2}[(\alpha_{xx} - \alpha_{yy})^2 + (\alpha_{yy} - \alpha_{zz})^2 + (\alpha_{zz} - \alpha_{xx})^2] + 3[(\alpha_{xy})^2 + (\alpha_{xz})^2 + (\alpha_{yz})^2].$$

The molecular polarizability anisotropy can be determined from measurements of the depolarization ratio of Rayleigh scattered light. The degree of depolarization ( $\sigma$ ) and Rayleigh scattering parameters (R) are expressed in terms of the mean polarizability and polarizability anisotropy as:

$$\sigma_n = \frac{6(\Delta\alpha)^2}{45(\overline{\alpha})^2 + 7(\Delta\alpha)^2}, \qquad \Re_n = 45(\overline{\alpha})^2 + 13(\Delta\alpha)^2$$

$$\sigma_p = \frac{3(\Delta\alpha)^2}{45(\overline{\alpha})^2 + 4(\Delta\alpha)^2}, \qquad \Re_{p\perp} = 45(\overline{\alpha})^2 + 7(\Delta\alpha)^2; \; \Re_{p\|} = 6(\Delta\alpha)^2$$

In the inelastic case of the Raman scattering in the same experimental setup the degree of depolarization is given by the same expressions but modified by derivatives with respect to the vibrational mode. The cross section for Raman scattering is completely determined by the scattering activity given by

$$A_p = 45(\overline{\alpha}')^2 + 7(\Delta\alpha')^2$$

where prime indicates a derivative with respect to the vibrational mode. The geometry optimization and all other calculations are done by Gaussian09 suite of program [18].

**Results:** Variation of structure and energetics:

In this work, we consider the lowest energy conformation of i-PrCN and its clusters with water molecules.In Figure 1, we showthe optimised structures of the isolated i-PrCN and its complexes with H2O. Some important geometric parameters (bond lengths and bond angles) are calculated by model (1) HF/631++ G(d,p), (2) MP2/6-31++G(d,p) and (3) DFT-B3LYP/6-31++G(d,p). The bond lengths and the bond angles calculated by the three models are similar. All the atoms are properly labelled and the inter-molecular H-bonds are represented by the dotted lines.

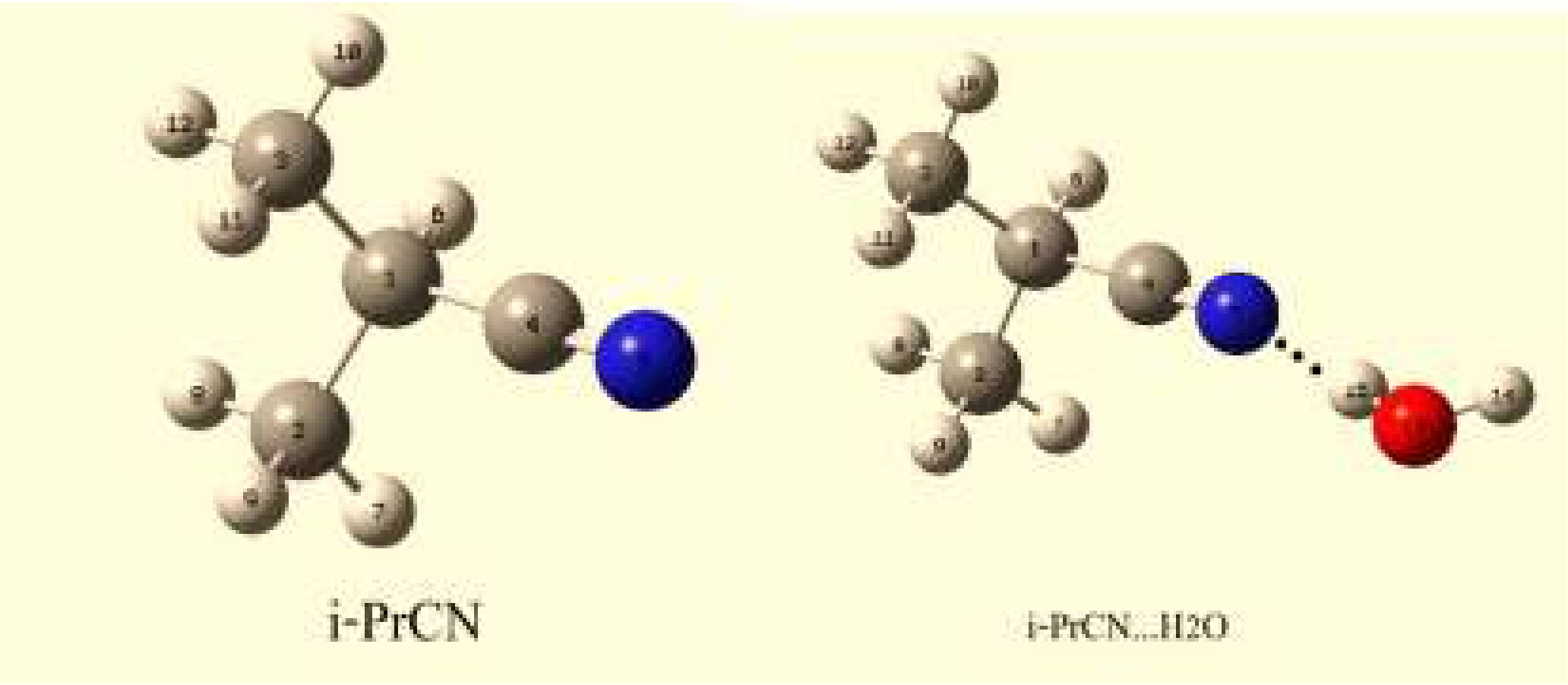

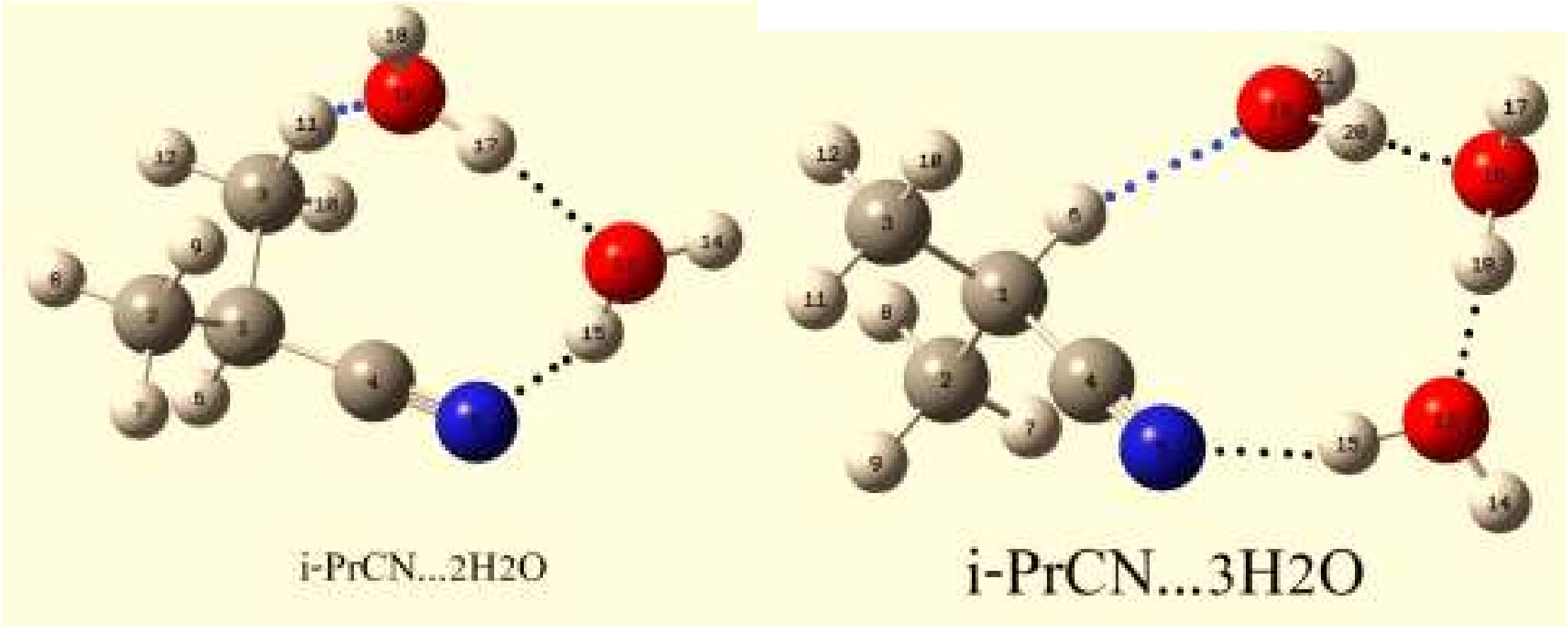

Fig 1. Isolated isopropyl cyanide (i-PrCN) and its complexes with water molecule (i-PrCN...H2O, iPrCN...2H2O and iPrCN...3H2O) optimized by model B3LYP/6-31+G(d,p).

Since all the models give similar optimized structures we present the structures obtained by DFT only. Some important geometric parameters (bond lengths and bond angles) calculated by model DFT-B3LYP/6-31++G(d,p) are shown in the figureitself. Here we observe that i-PrCN forms single conventional hydrogen bond (O-H...N) with one water molecule whereas trimer [i-PrCN-$(H2O)_2$] is stabilized by two conventional bonds (O-H...N, O-H...O) and one non-conventional H bond (C-H...O) and tetramer [i-PrCN-$(H2O)_3$]contains 3 conventional (O-H...N , O-H...O, O-H....O) and one non-conventional H bonds (C-H...O). All the H-bonds are found to have length ~ 2 $A^0$ . The C-H...O bonds O16···H11 and O19···H6 of trimer and tetramer respectivelyare found to have H-bond lengths slightly greater than the conventional H-bond lengths (for example, O16···H11 bond is greaterby 0.703 $A^0$ than the O13•••H17 bond in the trimer in DFT-B3LYP/6-31++G(d,p)). The angles of these C-H...O bonds are less than the conventional bonds (viz. O19···H6-C1 is less than ∠O16•••H20-O19by $23^0$ in the tetramer DFT-B3LYP/6-31++G(d,p)). These indicate that C-H...O bonds are weaker in strength compared to the conventional H-bonds.

Upon complexation, distances of C≡N, which acts as proton acceptor of all the H bonded complexes are slightly shorter than in isolated species except in [i-PrCN-$(H2O)_2$] case. As can be seen from the table 1, the N···H hydrogen bonding distance for tetramer is $1.0905A^0$ which is $0.977A^0$ and $0.892A^0$ shorter than those of dimer and trimmer respectively (calculated with DFT method). Other bonds (non H-bonds) didn't get affected much i.e. the lengths and angles remain almost same upon complexation.

**Table 1:** Selected geometries (bond lengths in $A^0$ and angles in degrees) of i-C3H7CN···$(H2O)_n$ (n=1=3) complexes computed with 6-31+G(d,p) basis set along with corresponding values of the isolated i-C3H7CN. (H bonds are represented by dotted lines).

| System | i-C3H7CN | | | i-C3H7CN···H2O | | | i-C3H7CN···2H2O | | | i-C3H7CN···3H2O | | |
|---|---|---|---|---|---|---|---|---|---|---|---|---|
| Method | HF | DFT | MP2 | HF | DFT | MP2 | HF | DFT | MP2 | HF | DFT | MP2 |
| C1-H6 | 1.0858 | 1.0982 | 1.0937 | 1.0857 | 1.0980 | 1.0935 | 1.0861 | 1.0983 | 1.0941 | 1.0827 | 1.0991 | 1.0916 |
| C3-H11 | 1.0846 | 1.0943 | 1.0888 | 1.0842 | 1.0941 | 1.0897 | 1.0831 | 1.0933 | 1.0886 | 1.0851 | 1.0942 | 1.0883 |
| C1-C4 | 1.4800 | 1.4729 | 1.4705 | 1.4790 | 1.4718 | 1.4696 | 1.4803 | 1.4707 | 1.4691 | 1.4788 | 1.4691 | 1.4677 |
| C4-N5 | 1.1364 | 1.1619 | 1.1827 | 1.1358 | 1.1606 | 1.1810 | 1.1366 | 1.1619 | 1.1818 | 1.1352 | 1.1616 | 1.1820 |
| ∠H12-C3-H11 | 108.3 | 108.5 | 108.7 | 108.4 | 108.4 | 108.6 | 109.0 | 109.1 | 109.3 | 105.2 | 108.6 | 108.6 |
| ∠C4--C1-H6 | 106.0 | 106.3 | 106.6 | 106.0 | 106.1 | 106.5 | 105.7 | 106.2 | 106.4 | 105.2 | 104.5 | 105.2 |
| ∠C1-C4-N5 | 179.7 | 179.5 | 179.2 | 179.6 | 179.3 | 179.0 | 179.5 | 198.6 | 179.0 | 179.4 | 178.9 | 179.5 |
| **N5···H15** | | | | 2.2366 | 2.0674 | 2.1007 | 2.1821 | 1.9829 | 2.0156 | 2.0807 | 1.0905 | 1.9444 |
| ∠O13-H15···N5 | | | | 174.6 | 179.6 | 178.9 | 155.5 | 157.7 | 153.7 | 163.5 | 165.0 | 162.5 |
| **O13···H17** | | | | | | | 1.9948 | 1.8531 | 1.8822 | 1.9320 | 1.7703 | 1.8078 |
| ∠O16-H17···O13 | | | | | | | 171.1 | 169.9 | 169.4 | 172.0 | 174.7 | 172.3 |
| **O16···H11** | | | | | | | 2.710 | 2.556 | 2.4733 | - | - | - |
| ∠O16···H11-C3 | | | | | | | 171.1 | 169.9 | 142.3 | - | - | - |
| **O16···H20** | | | | | | | | | | 1.954 | 1.8006 | 1.8379 |
| ∠O16···H20-O19 | | | | | | | | | | 178.2 | 179.2 | 178.8 |
| **O19···H6** | | | | | | | | | | 2.384 | 2.226 | 2.3157 |
| ∠O19···H6-C1 | | | | | | | | | | 155.5 | 156.2 | 130.8 |

The conventional hydrogen bonds of the type X–H..... Y (where X and Y are N or O) have been thoroughly studied in macromolecular structures from both experimental and theoretical point of view (1, 2, 7–9). On the other hand, close CH....O contacts which occur often in protein structures are also accepted as hydrogen bonds. It is found that weak CH.... O hydrogen bonds play a significant role in biological macromolecules' stabilization and function (3–6). Here we observe that i-PrCN forms single conventional hydrogen bond (O-H...N) with one water molecule whereas trimer [i-PrCN-$(H2O)_2$] is stabilized by two conventional bonds (O-H...N, O-H...O) and one non-conventional H bond (C-H...O) and tetramer [i-PrCN-$(H2O)_3$]contains 3 conventional (O-H...N ,O-H...O,O-H....O) and one non-conventional H bonds (C-H...O). All the H-bonds are found to have length ~2 $A^0$.The C-H...O bonds O16···H11and O19···H6of trimer and tetramer respectivelyare found to have H-bond lengths slightly greater than the conventional H-bond lengths (for example, O16···H11 bond is greater by 0.703 $A^0$ than the O13•••H17 bond in the trimer in DFT-B3LYP/6-31++G(d,p)). The angles of theseC-H...O bonds are less than the conventional bonds ( viz. O19···H6-C1 is lessthan ∠O16•••H20-O19by $23^0$ in the tetramerDFT-B3LYP/6-31++G(d,p)). These indicate that C-H...O bonds are weaker in strength compared to the conventional H-bonds.

Upon complexation, distances of C≡N, which acts as proton acceptor of all the H bonded complexes are slightly shorter than in isolated species except in [i-PrCN-$(H2O)_2$] case. As can be seen from the table 1, the N···H hydrogen bonding distance for tetramer is 1.0905$A^0$ which is 0.977$A^0$ and 0.892$A^0$ shorter than those of dimer and trimmer respectively (calculated with DFT method). Other bonds (non H-bonds) didn't get affected much i.e. the lengths and angles remained nearly same upon complexation.

**Table 2**

Total energies (E), binding energies without (ΔE) and with counterpoise corrections ($\Delta E^{CP}$) and with corrections for zero point vibrationalenergies($\Delta E^{CP+ZPE}$) of the iPrCN cluster.

| System | i-C3H7CN···H2O | | | i-C3H7CN···2H2O | | | i-C3H7CN···3H2O | | |
|---|---|---|---|---|---|---|---|---|---|
| Method | HF | DFT | MP2 | HF | DFT | MP2 | HF | DFT | MP2 |
| E (eV) | 286.0512 | 287.8433 | 286.9796 | 362.0949 | 364.2928 | 363.2309 | 438.1171 | 440.7442 | 439.4811 |
| ΔE (kcal/mole) | -4.3076 | -5.0644 | -5.4655 | -12.1002 | -14.7322 | -16.9219 | -20.7720 | -25.6402 | -27.6214 |
| $\Delta E^{CP}$ kcal/mole | -4.0358 | -4.6893 | -4.4968 | -11.0763 | -13.2138 | -12.8881 | -18.7820 | -22.7659 | -21.2440 |
| $\Delta E^{CP+ZPE}$ (kcal/mol) | -2.7175 | -3.2949 | -3.1112 | -7.2886 | -9.0953 | -8.7189 | -12.6880 | -16.1961 | -14.7385 |

Table 2 shows the energy of the dimer is lowest (287.8433 ev in DFT); it increases in trimer by 76.4945 ev (DFT) and is maximum for the tetramer (152.9009 ev more than dimer in DFT). Unlike the structural parameters, the energetic properties obtained by the three models differ appreciably from each other. Considering the binding energies (_*E*HB), as reported in Table 2,The HF values are markedly different from the other two values. It is expected as HF calculation doesn't take the electron correlation effect. But the other two methods, DFT and MP2 also differ appreciably from each other when we haven't taken the BSSE and zero point energy corrections. This difference is more when the number of water molecules increases. For example, the difference between binding energies calculated in DFT and MP2 methods is 0.4011 kcal/mol for dimer, 2.1897 kcal/mol for trimer and 1.9812 kcal/mol for tetramer. But when we take BSSE correction ($\Delta E^{CP}$)the difference reduces to 0.1925 kcal/mol for dimer,0.3257 kcal/mol for trimer and 1.5219 kcal/mol for tetramer. After the corrections for zero point vibrational energies($\Delta E^{CP+ZPE}$) the difference between two methods becomes 0.1837 kcal/mol for dimer, for trimer it is 0.3764 kcal/mol and for tetramer it is 1.4576 kcal/mol.

We also observe that i-PrCN···3H2O is more stable because the binding energy of the tetramer is least. It is more bound than dimer by 20.58 kcal/mole (DFT) and more bound than trimer by 10.91 kcal/mole (DFT).

**Table 3:** Dipole moments............ of i-C3H7CN···(H2O) $_n$ (n=1=3) complexes computed with 6-31+G(d,p) basis set along with corresponding values of the isolated i-C3H7CN.

| System | i-C3H7CN | | | i-C3H7CN···H2O | | | i-C3H7CN···2H2O | | | i-C3H7CN···3H2O | | |
|---|---|---|---|---|---|---|---|---|---|---|---|---|
| Method | HF | DFT | MP2 | HF | DFT | MP2 | HF | DFT | MP2 | HF | DFT | MP2 |
| $\mu$ | 4.3938 | 4.2178 | 4.0415 | 6.3877 | 6.2712 | 6.0349 | 1.9162 | 2.0070 | 1.816 | 1.6459 | 1.7714 | 1.6248 |
| $\bar{\alpha}$ | 46.608 | 50.951 | 48.656 | 53.483 | 59.558 | 56.686 | 60.202 | 68.358 | 65.072 | 67.531 | 77.804 | 73.611 |
| $\Delta\alpha$ | 16.419 | 18.916 | 17.389 | 21.295 | 26.023 | 23.367 | 17.582 | 21.287 | 19.873 | 19.905 | 25.267 | 23.311 |
| $\sigma_n$ | 0.0162 | 0.0179 | 0.0167 | 0.0206 | 0.0247 | 0.0221 | 0.0112 | 0.0127 | 0.0123 | 0.0114 | 0.0138 | 0.0132 |
| $\sigma_p$ | 0.0082 | 0.0091 | 0.0084 | 0.0104 | 0.0125 | 0.0112 | 0.0056 | 0.0064 | 0.0062 | 0.0057 | 0.0069 | 0.0066 |
| $\sigma_c$ | 0.0165 | 0.0183 | 0.0169 | 0.0210 | 0.0253 | 0.0225 | 0.0113 | 0.0129 | 0.0124 | 0.0115 | 0.0140 | 0.0133 |

In Table 3, we show the relevant parameters related to thevariation of electric properties of the single molecule i-C3H7CN and its cluster i-C3H7CN•••(H2O) n (n=1,2,3) calculatedby model (1), (2) and (3).The first row represents The dipole moment, μ (in D), average polarisability$\bar{\alpha}$ (in $\alpha^3$ ) represented by the second row, polarisability anisotropy Δα(in $\alpha^3$) and depolarisation ratio (σ) by third and fourth rows respectively. Here $\sigma_n$:degreeof depolarisation of natural light; $\sigma_p$:degree of depolarisation of plane-polarised light; $\sigma_c$ : degree of depolarisation of circularly polarised light.

As we can see, the presence of a neighbouring molecule has a significant effect on the values of dipole moments and polarizabilities .The formation of hydrogen bond is found to have significant effect on calculation of dipole moments and polarizabilities.The dipole moment gets increased by 2.0534 D (DFT) when i-PrCN forms dimer with one water molecule but decreases by 2.2108 D (DFT) in trimer and even further reduced by 2.4464 D (DFT) in tetramer. Thus, the dipole moment increases almost additively in this particular case of linear H-bonded structure which is the dimer, whereas, for trimer and tetramer, the dipole moments reduce significantly because of their symmetric geometry. On the other hand, polarizability varies in a more predictable way. The average polarisability values differ appreciably from each other in the three methods. Its value is gradually increased when i-PrCN forms clusters with water molecules. It is higher by 8.607 (DFT) for dimer than a single molecule, by 17.407 (DFT) for trimer and by 26.853 (DFT) for tetramer.The polarizability anisotropy, however, shows different variation patterns for different clusters. It is found to be maximum for dimer(28.32 a.u.) and minimum for trimer (13.90 a.u.) at theDFT-B3LYP/6-31++G(d,p)level of calculation. The polarisability anisotropy is increased by 7.107(DFT) in dimer than isolated i-PrCNmolecule , in trimer by 2.371(DFT) and by 6.351(DFT) for tetramer. In Table 3, we present the calculated values of degree of depolarization (or depolarization ratio) of light (r) which is, in fact, one of the parameters most frequently observed in the light scattering experiments.[47] The changes in polarizability and anisotropy of the polarizability because of H-bond formation cause the variation in degree of depolarization in the hydrogen-bonded clusters. All the components of depolarization of light, both natural and polarized (both plane-polarized and circularly polarized) for the isolated molecule as well as the clusters formed by these molecules, calculated at all the threetheoretical levels are presented. In Table 3, we observe different kinds of variations in the degree of polarization. These occur due to different variations in the polarization and anisotropy due to the formation of H-bonds. For example, $\sigma_n$ in dimer is greater by .0068; in trimer it is lower by .0052 and for tetramer lesser by .0041 in DFT method. As far as the variations of degree of polarization due to polarized light (rp and rc) are concerned, we observe, more or less, the same behavior as that of $\sigma_n$ .

**Table 4:** Rayleigh parameters of i-C3H7CN···(H2O) n (n=1=3) complexes computed with 6-31+G(d,p) basis set along with corresponding values of the isolated i-C3H7CN.

| System | i-C3H7CN | | | i-C3H7CN···H2O | | | i-C3H7CN···2H2O | | | i-C3H7CN···3H2O | | |
|---|---|---|---|---|---|---|---|---|---|---|---|---|
| Method | HF | DFT | MP2 | HF | DFT | MP2 | HF | DFT | MP2 | HF | DFT | MP2 |
| $R_n$ | 101259.9 | 121473.7 | 110465.6 | 134614.9 | 168427.5 | 151697.1 | 167109.3 | 216167.7 | 195678.7 | 210370.3 | 280707.7 | 250898.5 |
| $R_{p\perp}$ | 99642.3 | 119326.6 | 108651.4 | 131893.9 | 164364.2 | 148420.8 | 165254.6 | 213448.7 | 193309.1 | 207993.1 | 276877.2 | 247637.9 |
| $R_{p\parallel}$ | 1617.5 | 2147.1 | 1814.2 | 2721.0 | 4063.2 | 3276.2 | 1854.6 | 2718.8 | 2369.6 | 2377.2 | 3830.6 | 3260.5 |

In Table 4, we present the Rayleigh light scattering parameters of the isolated molecule as well as its hydrogen-bonded structures, calculated by the use of the three different models as aforementioned for both natural and parallel and perpendicular polarized light. As can be observed from the table, the formation of the hydrogen bonding among the molecules increased significantly the Rayleigh intensities for all components of light. However, the proportional increase in Rn practically remains equal in all three basis sets for all the molecules. The Rayleigh parameters for the perpendicular-polarized light show more or less the same behaviour. However, the Rayleigh parameters for the parallel-polarized light do not present the same pattern. It increases in dimer by 1916.1but in trimer the increase is significantly less; viz. in trimer the increase is only 571.7 and in tetramer it increases by 1683.5.

It is clear from table 4 thatthe formation of H-bond leads to considerable change inthe Rayleigh parameters.Since H-bond formation, in general,introduces change in the anisotropy, it leads to a correspondingchange in the Rayleigh intensities and depolarisationratios.Rayleigh light scattering is related to dipole polarisability which is, in fact, the basic property that describesthe interaction of a molecule with an electric field.The anisotropy of the electric dipole polarisability is an important parameter in order to study themolecular interactionwith applied electric field and to understand the long-rangeintermolecular induction and dispersion forces.

**Conclusions:** A theoretical study of the effect of hydrogen bond formation on structure, energetic and electric properties related to light scattering of microhydrated isopropyl cyanide [(i-PrCN-(H2O)n) (n=1-3)] have been explored by using three different quantum chemical models (i) HF/ 6-31++G(d,p), (ii) MP2/6-31++G(d,p) and (iii) DFT B3LYP/6-31++G(d,p). All the models give similar optimized structures and it is noted that the geometries remain same upon complexation. Unlike the structural parameters, the energetic properties obtained by the three models differ appreciably from each other. Presence of water molecule has a significant effect on the values of dipole moments and polarizabilities. Rayleigh intensities which are calculated using mean polarizibility and polarizibility anisotropy are increased much due to the formation of hydrogen bonding.